\documentclass[]{aa}
\usepackage{graphicx}
\usepackage{psfig}
\usepackage{epsfig}
\begin{document}

   \thesaurus{02.19.2;
              03.13.2;
              08.02.2 HD 100546}

   \title{ADONIS observations of the HD 100546 circumstellar dust disk 
   \footnote{based on observations
   collected on the ESO 3.6m telescope at La Silla, proposal 63.H-0239}}

   \subtitle{}
   \mail{epantin@cea.fr}
   \author{E. Pantin\inst{1}, C. Waelkens\inst{2}, and P.O. Lagage\inst{1}}
  
   \offprints{E. Pantin}

   \institute{DSM/DAPNIA/Service d'Astrophysique, CEA/Saclay, F-91191
    Gif-sur-Yvette, France \and Instituut voor Sterrenkunde, Celestijnenlaan 
    200B, B-3001 Leuven, Belgium}

   \date{Received; accepted}
 
   \titlerunning{ADONIS observations of HD 100546}
   \authorrunning{E. Pantin et al.}      
   \maketitle

   \begin{abstract}
    We report in this letter the first resolved images of the circumstellar
    dust disk around the Pre-Main-Sequence star HD 100546. 
    These near-infrared images were obtained in J and
    short K bands with the adaptive optics
    system ADONIS at the ESO observatory. A bright disk extending 
    2 arcsec (200 AU) far 
    from the star and viewed with an inclination with respect to the line 
    of sight around 
    50 degrees is revealed by the observations. Using a simplified model of
    light scattering, we find that the dust disk density peaks at a 
    distance of 40 AU from the star, and the FWHM of this dense ring is 
    typically 20 AU. This type of disk is believed to be a denser precursor of 
    $\beta$ Pictoris-like
    main-sequence disks. The structure of the disk is compared with 
    the best known example of the class of ``debris disks'', 
    the disk surrounding $\beta$ Pictoris.

   \keywords{Stars: individual: HD 100546, Methods: data analysis, Scattering}
   \end{abstract}

%

\section{Introduction}

The discovery of infrared excesses around an important
fraction of main-sequence stars (Aumann et al. 1984; Plets \& Vynckier 1999)
and attributed to the presence of cool dust grains orbiting the star 
and geometrically arranged in 
a disk, has triggered a lot of studies
because these disks may be linked to planetary formation. 
For a long time the disk around the star $\beta$ Pictoris was the only example
to be resolved both in the visible (Smith \& Terrile 1984) and
in the mid-infrared range (Lagage \& Pantin 1994).
Recent discoveries of young ``debris'' circumstellar disks around 
relatively ``old'' and isolated (i.e. not associated to any 
star forming region) stars have shown that this phenomenon 
extends towards Pre-Main-Sequence stars (cf the photometric 
survey by Malfait et al. 1998a). 
The detection and resolution of disk around stars such as HD 141569 
(Weinberger et al. 1999; Augereau et al. 1999) or HR 4796A thanks to 
high-resolution observations in the visible/near-infrared 
range (Schneider et al. 1999; Augereau et al. 1999) 
or mid-infrared images (Koerner et al. 1998; 
Jayawardhana et al. 1998) have shown the possibility  
to observe and study precursors to main-sequence dust disks.
These so-called ``baby-$\beta$ Pic''
dust disks are the denser precursors to main-sequence debris disks. 
IRAS and ISO/SWS observations have shown 
that they usually produce a huge infrared excess (typically 250 times the 
infrared excess produced by the $\beta$ Pic disk); some 
of them, as HD 100546 for instance, showing prominent signatures
of crystalline water ice (Waelkens et al. 1996, Malfait et al. 1998b), 
are particularly interesting
targets in the visible/near-infrared because of high particle albedo.
Observing this class of disks at various stages of evolution 
will help in finding a comprehensive scenario for the origin, the evolution,
and the lifetime of the Vega phenomenon, i.e. how these disks form, 
how long they last, and how they disappear.\\
We report in this paper the first images of the
disk around HD 100546, an isolated young main-sequence star.
Using a model of scattering 
by dust grains, we derive the morphology of the disk outwards 
10 AU and show evidence for a density maximum around 40 AU. 
The structure of the disk is compared to the best known example 
of the Vega Phenomenon, the disk around $\beta$ Pic. 
  

\section{Observations}

We observed the HD 100546 star (m$_{\mbox{\tiny{V}}}$=6.7) using the SHARP II+ camera 
coupled with the adaptative optics system
ADONIS, and mounted on the ESO 3.6m telescope at La Silla, Chile.
A pre-focal optics coronograph (Beuzit et al. 1997) was used to 
reject the direct starlight and increase the
integration time in each elementary exposure. 
The mask size we used has a diameter of 0.74\arcsec. 
J (1.25 $\mu$m) and Ks (2.15 $\mu$m) 
band exposures were obtained on nights 23 and 
24 of June, 1999. The seeing was quite variable the first night, but had 
decreased to a value of $0.7''$ at the time of the J and Ks observations.
It was stable at a value around $0.85''$ during the second night. 
We spent a total observing time of 1550 s in the Ks band and 2600 s in the J band.
The Point Spread Function (PSF) was frequently monitored thanks 
to interlaced observations of two reference stars, HD~97218
and HD~101713. These reference stars were used later in the reduction 
process to remove the wings of the stellar PSF and for photometric calibration.
Observations of empty fields were also performed to estimate and 
remove the background flux which can be important in the Ks band.

\section{Reduction procedure}

\begin{figure}
\includegraphics[scale=0.40,keepaspectratio=false,
clip=true]{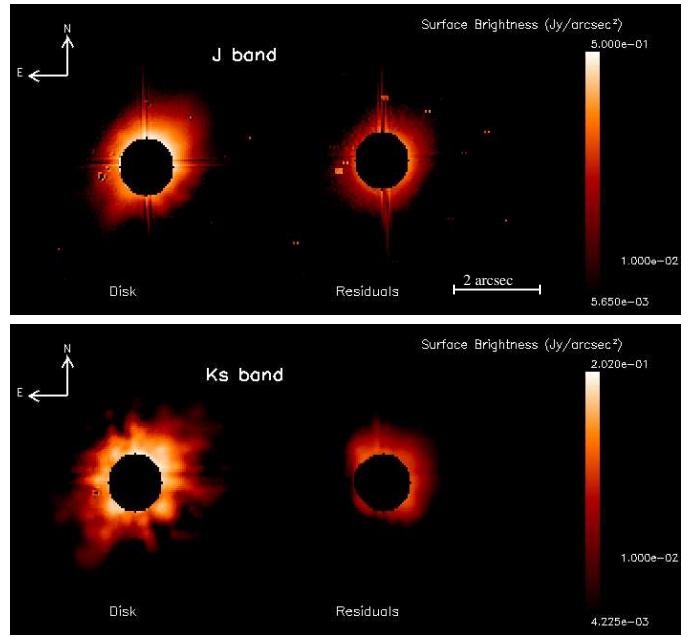}
\caption{Upper left : the disk seen in the J band, 
North is to the top, East to 
the left. The pixel scale is 0.035 arcsec/pixel. Upper right, the corresponding 
residuals obtained when applying the reduction method described in the 
text to two reference stars. 
Lower frame : same as upper frame but in the Ks filter. The ``cross-like'' pattern
(with fingers orientations at 45,135,225, and 315 degrees)
superimposed to the disk emission is produced by astigmatism residuals not
corrected by the adaptive optics loop, showing that, contrarily to J band 
where the correction residuals are still dominated by the seeing, 
in Ks band under good conditions, one can reach high Strehl ratio values (i.e. 
system residuals dominating the uncorrected part of the seeing)}
\label{Figure1}

\end{figure}

First, standard reduction techniques were applied
to the data : bias subtraction, flat-field correction. 
For each filter, we obtained a set of HD 100546 observations and
corresponding PSFs. In spite of the use of a coronograph mask,  
the disk surface brightness is still dominated by the
stellar emission at any distance from the star.
In order to retrieve the disk image, one has to remove numerically the starlight wings. 
The rough substraction of a scaled PSF to HD 100546 images gives poor
results because of slight shifts in position on the array 
between the reference and the object, uncertainties on the fluxes (given by 
the literature), and a residual 
background (ADONIS bench emission, different airmasses). 
We developed a specific method to get an optimum subtraction. 
For each couple of HD~100546 image ($\mbox{Obj}$) and corresponding $\mbox{Psf}$, 
we have to find 3 parameters :  a shift ($\delta x$,$\delta y$)
between the two images, a scaling factor $R$, and a residual background 
$Bg$. These parameters are estimated by minimizing the following error 
functional :
\begin{equation}
 J = \sum \arrowvert \mbox{Obj} - S(\mbox{Psf}/R,\delta x, \delta y)- Bg  \arrowvert 
\end{equation}
where the S function stands for image shift. The sum is performed on a 
set of pixels (typically 1000) located in a region of the images where 
no disk emission is expected, and from which non accurate ones 
are excluded (bad pixels, pixels belonging to 
area contaminated by diffracted light from the coronograph support). 
The functional minimum is found using a 0-order minimization 
algorithm called Powell method (Press 1996).
The method was first checked on simulated data whose input 
parameters (shifts, scaling factor) were recovered with an accuracy better
than 5 \%. 
In order to evaluate the errors
in the subtraction process, test the stability of the PSF,
and make sure that the disk obtained is not an artifact created
by the process, the same substraction process was applied to 
the two different PSFs. Figure \ref{Figure1} shows 
the resulting images in the J and Ks filters of the disk 
and the corresponding residuals between the two reference stars.
The position of the star under the mask, and therefore
the disk center is also carefully determined in this procedure 
thanks to offset observations (out of the coronograph mask) 
of the reference stars and the object.

\section{Results}
Since the dataset in J band is more accurate (because of a larger 
integration time, better observing conditions, and a better 
sensitivity of the whole system) 
than in Ks band,
the geometrical parameters and the disk model are based on the J band image. 
We checked in a second step that the results are compatible with the Ks band data.

\subsection{Disk geometrical parameters}
Given a disk inclination with respect to the line of sight which 
is $\approx$ 45$^{\circ}$, we expect to observe some deviations
from the perfect ellipse case. Indeed, particles relatively large with respect 
to the wavelength produce non-isotropic scattering. 
One simple way to reduce this effect is to 
symmetrize the disk i.e. construct an isotropic-like virtual disk
by averaging the original disk image and its symmetrical image 
with respect to the major axis.   
Best ellipse fitting on this symmetrized images gives the following results:
\begin{itemize}
\item{} disk position angle on the sky of $37  \pm 5 $ degrees with respect to east direction.
\item{} ellipse major to minor axis ratio of 1.2 implying a disk inclination of 
   $50 \pm 5$ degrees with respect to the line of sight (or from the 
   edge-on case).
\item{} disk extension : $\approx$ 2.0 arcsec from the star 
   corresponding to a distance to the 
   star around 200 AU (assuming a distance to the Earth of 103 parsecs 
   (van den Ancker et al. 1998) and a limiting sensitivity of 5 mJy/arcsec$^2$.
\end{itemize}

\subsection{Photometry}

Using the stars HD 101713 (B9V) and HD 97218 (G8) as references whose V magnitude
is known and using appropriate V-J and V-Ks color indices, we derived the following photometry
for the disk in J and Ks band: a disk flux in J band of $0.32 \pm 0.06$ Jy outside the 
coronograph mask and a maximum surface brightness of $0.5 \pm 0.1$ Jy/arcsec$^2$.
From images of HD 100546 and corresponding reference star images (HD 101713
   and HD97218) obtained when shifting the targets 
   outside of the mask, 
   we can deduce a total of scattered flux in the J band around 1 Jy; note 
   that we get also a disk-shaped emission when removing a scaled
   reference to the HD 100546 image obtained outside the mask, but its
   signal to noise is very low.
   This result is compatible with the infrared excess computed considering a 
   Kurucz model of the star (B9V spectral type). 
   Consequently, the disk flux outside the coronograph mask represents 
   roughly 30\% of the total disk emission. \\  
In the Ks band, we find an emission of $0.25 \pm 0.1$ Jy outside the 
   coronograph mask and a maximum surface brightness of $0.25 \pm 0.1$
   Jy/arcsec$^2$

\subsection{Model and observation fit}

\begin{figure*}

\resizebox{12cm}{!}{\includegraphics[scale=0.7,clip=true]{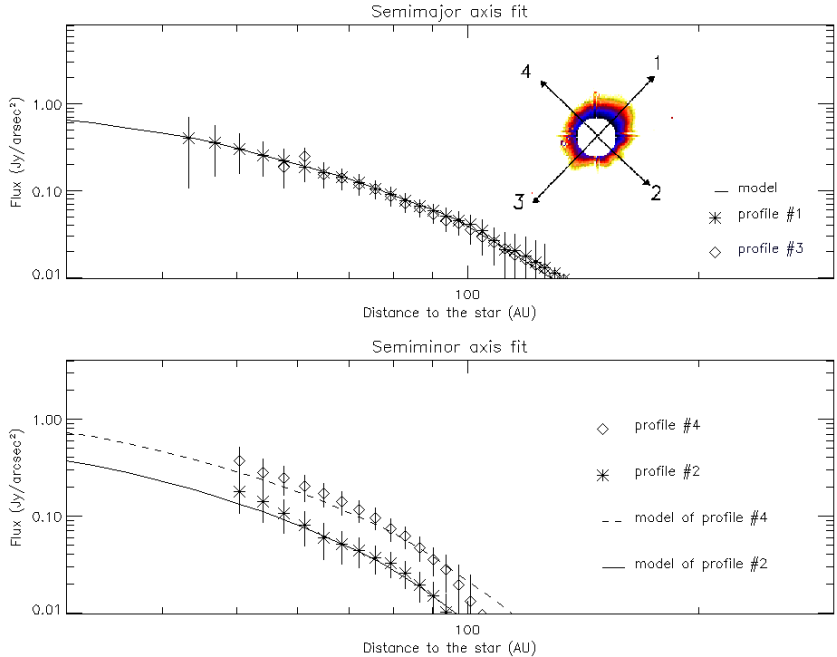}}
\hfill
\parbox[b]{55mm}{

\caption{Fit of the data profiles. Upper plot, full lines : the 
 modeled profiles, stars
 and diamonds represent the data profiles 1 and 3 defined on the disk image
 inserted. Lower plot, dashed line : model of profile 4 (dominant 
  forward scattering), plain line : model of profile 2. As seen on the lower plot, the
  scattering asymmetry is well reproduced by an anisotropic phase function (Henyey-Greenstein, 
  see the text) with $g=0.2$, corresponding to a particle size around 0.1 $\mu$m.}

\vspace{2cm}

\label{fig_fit}

}

\end{figure*}

In order to derive physical parameters of the disk, we built a numerical
model of starlight scattering by an optically thin disk made of dust particles. 
Assuming that no multiple scattering is occurring in the disk, and since the exact 
physical properties of the particles are poorly known 
(composition, shape, size distribution),
we chose to use a global, single-parameter, scattering phase function. The most
practical one, although not realistic, is the Henyey-Greenstein 
phase function (Henyey \& Greenstein 1941). 
The disk is assumed to be seen with an inclination derived above, and to have 
a proper thickness similar to the one of $\beta$ Pic parameterized following Artymowicz 
et al. 1989. The disk midplane density is described
by a scattering area which follows a series of broken power laws. We start with
a first guess deduced from the radial surface brightness, and we iterate
on the parameters until we obtain a satisfactory fit along the four directions
defined on figure \ref{fig_fit}.  

\begin{figure}

\includegraphics[scale=0.53,clip=true]{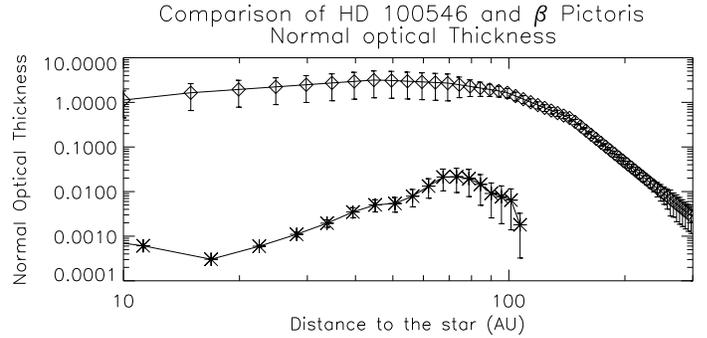}

\caption{The normal optical thickness of the disk deduced from model fitting
         of the data (diamonds). 
         We have overplotted (stars) the normal optical 
         thickness found in the case of the disk around $\beta$ Pictoris 
         (Pantin, Lagage \& Artymowicz 1998)
         }
\label{fig_taus}

\end{figure}

\subsection{Fit results}
The best fit model provides the following parameters (one must keep in mind
that they are model-dependent) :
\begin{itemize}

\item{} The Henyey-Greenstein phase function parameter found is $g=0.2$ corresponding to
         dominant spherical particles with size around 0.1 micron.
\item{}  A radial normal optical thickness fitted by a series of broken power laws 
         whose exponents are : 0.6 in the 
         range [10,45]AU, $-$0.4 in the range [45,70]AU, 
         $-$1.35 in the range [70,100]AU,
         $-$5 outwards from 100 AU, see figure \ref{fig_taus}.
             
\item{}  An exponentially decreasing vertical profile, and an opening angle
         around 0.1 radian.

\item{} A total scattering area of 10$^{29}$ cm$^2$ and a dust mass of 0.02 M$_{\oplus}$.

\item{} An approximate disk surface density at maximum around $3 \times 10^{14}$
        particles/m$^2$ assuming single-sized particles of 0.1 $\mu$m, 
        or an equivalent
        maximum scattering area of the order of the unity/m$^2$.

\end{itemize}

\section{Discussion}
The radial surface brightness distribution shows that
the disk density peaks at a distance of 40 AU from the star corresponding
to the edge of the planetary zone in our solar system.
The normal optical thickness found is more than 100 times 
larger than in the case of the $\beta$ Pic disk. Simultaneously, the presence of 
sub-micronic particles sizes,
well below the radiation pressure cut-off in sizes, suggests that 
huge quantities of very small particles are produced continuously 
by evaporation or collisions of planetesimals.  
We are probably witnessing 
a analogue of the young Kuiper belt with planetesimals undergoing 
violent agitation. If the dust particles are the result of collisions, 
a high rate is required to produce the
observed disk. 
This implies that gravitational perturbations occur frequently    
in this disk, maybe because of giant planets. 
If the disk is replenished by planetesimal evaporation as 
suggested by Lecavelier des Etangs (1998) in the case of the $\beta$ Pic disk,
10$^9$ planetesimals (40 km size) whose orbits were recently perturbed by planet migration 
are required to maintain the disk. Some of them which orbits are gravitationally perturbed
might be the origin of the casual redshifted absorption profiles observed in some
of the spectra of HD 100546 (Grady et al. 1997), a phenomenon highly similar to what is observed
in the system of $\beta$ Pictoris (Lagrange-Henry et al. 1987, Beust et al. 1996).

\section{Conclusion}
Thanks to coronograph adaptive optics observations we could detect 
the disk surrounding the PMS star HD 100546. This detection of a disk
which is believed to represent a class of precursors to $\beta$ Pic-like
disks adds a new clue in the understanding of the formation and evolution
of these debris disks. 
Other imaging observations of disks at various stage of evolution 
are necessary to sketch more closely the possible scenario 
of such a disk creation and evolution. This should be achieved by next generation
instruments such as the VLT/NAOS (adaptive optics) and 
VLT/VISIR (mid-infrared) instruments or the NGST.

\begin{acknowledgements}
  
We would like to express our warm gratitude to the
persons who helped in achieving these observations : D. Le Mignant, F. Marchis,
O. Marco, V. Meri$\tilde{n}$o, E. Wenderoth, and the whole 3.6m team.

\end{acknowledgements}

\end{document}